\documentclass[twocolumn]{svjour3}
\smartqed
\usepackage[T1]{fontenc}
\usepackage{graphicx}
\usepackage{subfigure}
\usepackage{cite}
\usepackage{tipa}
\usepackage{url}
\usepackage{float}
\usepackage{amsmath,amssymb,amsfonts}
\usepackage{algorithmic}
\usepackage{graphicx}
\usepackage{textcomp}
\usepackage{adjustbox}
\usepackage[graphicx]{realboxes}
\usepackage{rotating}
\usepackage[misc]{ifsym}
\usepackage{setspace}
\usepackage{algorithm}
\usepackage{algorithmic}
\usepackage{footnote}

\begin{document}

\title{COVID-19 Detection Based on Self-Supervised Transfer Learning Using Chest X-Ray Images}

\author{Guang Li $^{1}$ \and Ren Togo $^{2}$ \and Takahiro Ogawa $^{2}$ \and Miki Haseyama $^{2}$
}

\institute{Guang Li \\
           guang@lmd.ist.hokudai.ac.jp \\ \\
           Ren Togo \\
           togo@lmd.ist.hokudai.ac.jp \\ \\
           Takahiro Ogawa \\
           ogawa@lmd.ist.hokudai.ac.jp \\ \\
           Miki Haseyama \\
           mhaseyama@lmd.ist.hokudai.ac.jp $\textrm{\Letter}$ \\ \\
           $^{1}$ 
              Graduate School of Information Science and Technology, Hokkaido University, Sapporo, Japan \\ \\
           $^{2}$
              Faculty of Information Science and Technology, Hokkaido University, Sapporo, Japan 
}

\date{Received: date / Accepted: date}

\maketitle

\begin{abstract}
{\it Purpose:~} 
Considering several patients screened due to COVID-19 pandemic, computer-aided detection has strong potential in assisting clinical workflow efficiency and reducing the incidence of infections among radiologists and healthcare providers. 
Since many confirmed COVID-19 cases present radiological findings of pneumonia, radiologic examinations can be useful for fast detection. Therefore, chest radiography can be used to fast screen COVID-19 during the patient triage, thereby determining the priority of patient’s care to help saturated medical facilities in a pandemic situation.

{\it Methods:~} In this paper, we propose a new learning scheme called self-supervised transfer learning for detecting COVID-19 from chest X-ray (CXR) images. We compared six self-supervised learning (SSL) methods (Cross, BYOL, SimSiam, SimCLR, PIRL-jigsaw, and PIRL-rotation) with the proposed method. Additionally, we compared six pretrained DCNNs (ResNet18, ResNet50, ResNet101, CheXNet, DenseNet201, and InceptionV3) with the proposed method. We provide quantitative evaluation on the largest open COVID-19 CXR dataset and qualitative results for visual inspection.

{\it Results:~} Our method achieved a harmonic mean (HM) score of 0.985, AUC of 0.999, and four-class accuracy of 0.953. We also used the visualization technique Grad-CAM++ to generate visual explanations of different classes of CXR images with the proposed method to increase the interpretability.

{\it Conclusions:~}  Our method shows that the knowledge learned from natural images using transfer learning is beneficial for SSL of the CXR images and boosts the performance of representation learning for COVID-19 detection. Our method promises to reduce the incidence of infections among radiologists and healthcare providers.
\end{abstract}
\section{Introduction}
\label{sec1}
The Coronavirus disease 2019 (COVID-19) caused by severe acute respiratory syndrome coronavirus 2 (SARS-CoV2) has emerged as one of the deadliest viruses of the century, resulting in about 326 million infected people and over 5.5 million deaths worldwide as of January 17, 2022.\footnote{https://covid19.who.int}
Despite the unprecedented COVID-19 pandemic, medical facilities have faced many challenges, including a critical shortage of medical resources, and many healthcare providers have themselves been infected~\cite{leung2020covid}.
Due to the highly contagious nature of COVID-19, early screening has become increasingly important to prevent its further spread and reduce the incidence of infections among radiologists and healthcare providers~\cite{whiteside2020redesigning}.
\par
As a sequel to that, polymerase chain reaction is currently considered the gold standard in COVID-19 confirmation because of its high accuracy and takes several hours to get the result~\cite{tahamtan2020real}.
Since many confirmed COVID-19 cases present radiological findings of pneumonia, radiologic examinations can be useful for fast detection~\cite{qi2021chest}.
Therefore, chest X-ray (CXR) images can be used to fast screen COVID-19 during the patient triage, thereby determining the priority of patient’s care to help saturated medical facilities in a pandemic situation~\cite{quiroz2021identification}.
\par
Based on the above findings, several studies have been conducted to detect COVID-19 using CXR images with the help of different deep learning technologies~\cite{cau2022long}. 
Different neural network architectures, transfer learning techniques, and ensemble methods have been proposed to improve the performance of automatic COVID-19 detection~\cite{ozturk2020automated, khan2021covid, khan2020coronet}.
For example,~\cite{minaee2020deep} proposed to use transfer learning with several deep convolutional neural networks (DCNNs), such as VGGNet~\cite{simonyan2014very}, ResNet~\cite{he2016deep}, and DenseNet~\cite{huang2017densely} to detect COVID-19 using CXR images.
Their results achieved good performance on a small COVID-19 dataset.
Moreover,~\cite{ismael2021deep} proposed to ensemble deep feature extraction for the support vector machines classifier, fine-tuning of CNN, and end-to-end training of CNN to obtain better COVID-19 detection performance.
\par
Recently, self-supervised learning (SSL) methods have received widespread attention~\cite{jing2020self}.
Unlike supervised learning, self-supervised learning can learn good representations without manually designed labels, reducing the labeled cost and time.
For example,~\cite{noroozi2016unsupervised} and~\cite{gidaris2018unsupervised} proposed studies that played a jigsaw game on images and predicted the rotation degrees of images for learning good representations, respectively.
Furthermore, SSL methods have been effective on different medical datasets~\cite{zhou2021models}.
Transfer learning~\cite{pan2009survey} is a technique where a model trained on one task is reused on a second related task, which is commonly used in medical image analysis.
Since the learned representations using SSL on the target dataset are insufficient, transfer learning from different datasets may make up for the shortcomings of SSL to obtain better representations.
Several studies on different tasks (e.g., forecasting adverse surgical events~\cite{chen2021forecasting} and hand mesh recovery~\cite{chen2021self}) have shown the effectiveness of combining SSL and transfer learning.
\par
In this paper, we propose a new learning scheme called self-supervised transfer learning for detecting COVID-19 using CXR images.
Our method consists of three stages, first is the supervised pre-training on labeled natural images.
The next stage is the self-supervised pre-training on unlabeled CXR images, and finally, the fine-tuning on labeled CXR images.
We show that the knowledge learned from natural images with transfer learning is beneficial for SSL on the CXR images and boosts representation learning performance for COVID-19 detection.
Our method can learn discriminative representations from CXR images.
Furthermore, We realized promising detection results even when using a few labeled data for fine-tuning.
\begin{figure*}[t]
        \centering
        \includegraphics[width=16cm]{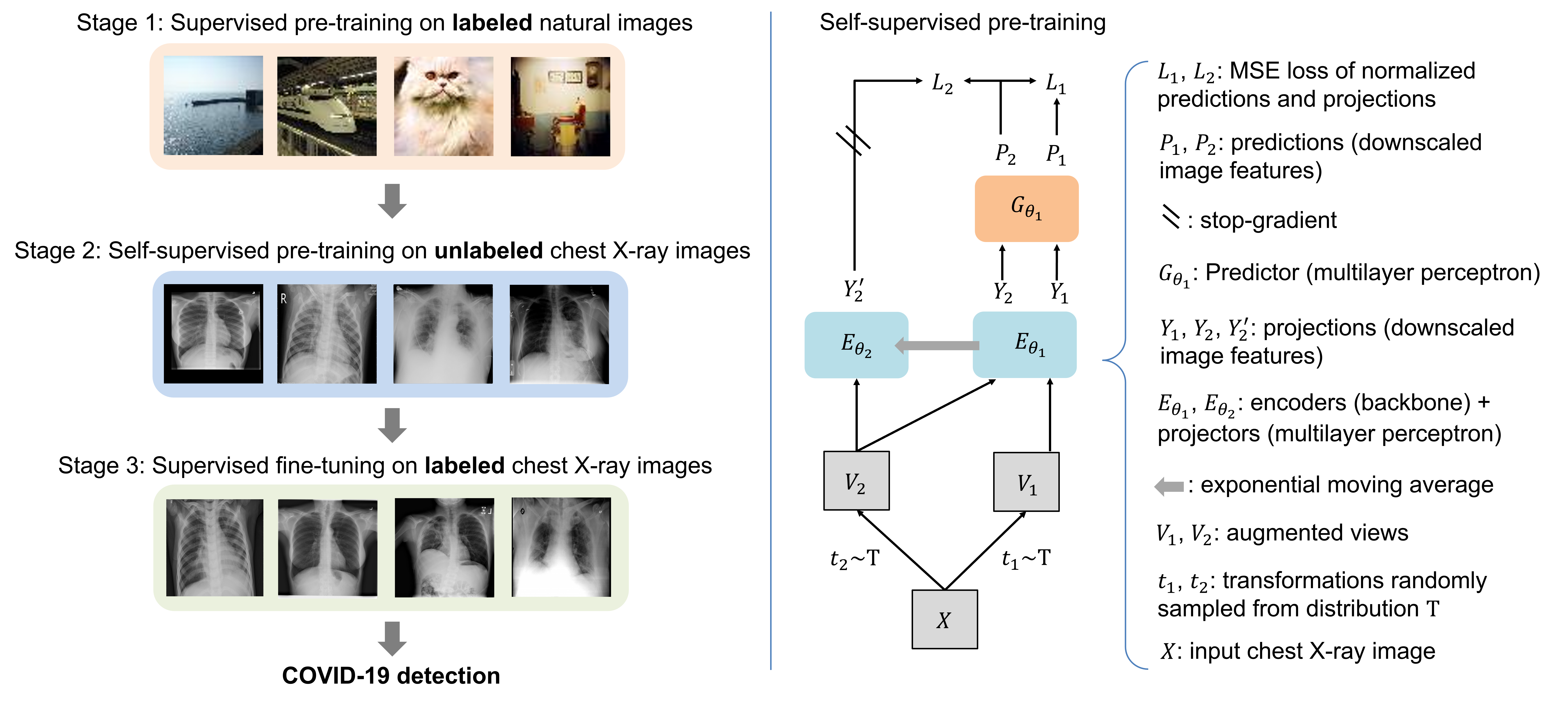}
        \caption{Overview of the proposed method. The left shows the concept illustration of our method and the right shows the self-supervised pre-training process of our method.}
        \label{fig1}
\end{figure*}
\section{Theoretical background}
\label{sec2}
Considering that the learned representations using SSL on the target dataset are insufficient, transfer learning from different datasets can be employed to augment shortcomings of SSL and obtain better representations.
We propose to learn discriminative representations from CXR images by using transfer learning and SSL.
We show that the knowledge learned from natural images using transfer learning benefits the SSL on the CXR images and boosts representation learning performance for COVID-19 detection.
\par
An overview of the proposed method is shown in Fig.~\ref{fig1}.
Our method consists of three stages, first is supervised pre-training on labeled natural images, next is self-supervised pre-training on unlabeled CXR images, and final is supervised fine-tuning on labeled CXR images.
Specifically, we use the model pre-trained on ImageNet~\cite{deng2009imagenet} as the encoder in self-supervised pre-learning.
Then, after self-supervised pre-training on unlabeled CXR images, a well-trained encoder is obtained, and a classifier is added behind it for supervised fine-tuning.
Finally, after the supervised fine-tuning on labeled CXR images, we can use it to detect COVID-19 effectively.
\par
As shown in Fig.~\ref{fig1}, given an input CXR image $X$ without label information, two transformations $t_{1}$ and $t_{2}$ are randomly sampled from a distribution $T$ to generate two views $V_{1} = t_{1}(X)$ and $V_{2} = t_{2}(X)$~\cite{he2020momentum}.
Specifically, these transformations are combined using standard data augmentation methods such as cropping, resizing, flipping, and Gaussian blurring~\cite{li2022tri}.
Thus, $Y_{1}$, $Y_{2}$ and $Y'_{2}$ are output representations processed by the encoders $E_{\theta_{1}}$ and $E_{\theta_{2}}$.
$P_{1}$ and $P_{2}$ are output representations processed by the predictor $G_{\theta_{1}}$.
The predictor $G_{\theta_{1}}$ is designed to make the network structure asymmetric, thereby preventing a collapse in learning~\cite{grill2020bootstrap}.
Finally, we define the loss $L$ for updating parameters as follows:
\begin{equation}
\label{eq1}
L = L_{1} + L_{2},
\end{equation}
where $L_{1}$ and $L_{2}$ are used to compare the normalized representations from two views of the same image as follows:
\begin{equation}
\label{eq2}
L_{1} 
= || \hat{P}_{1} - \hat{P}_{2} ||_{2}^{2} = 2 - 2\cdot\frac{\left \langle P_{1},P_{2} \right \rangle}{ || P_{1} ||_{2} \cdot || P_{2} ||_{2}},
\end{equation}
\begin{equation}
\label{eq3}
L_{2}
= || \hat{P}_{2} - \hat{Y}'_{2} ||_{2}^{2} = 2 - 2\cdot\frac{\left \langle P_{2},Y'_{2} \right \rangle}{ || P_{2} ||_{2} \cdot || Y'_{2} ||_{2}},
\end{equation}
where $\hat{P}_{i} = P_{i}/ || P_{i} ||_{2}$ and $\hat{Y}'_{i} = Y'_{i}/ || Y'_{i} ||_{2}$ represent the normalized representations of $V_{i}$ ($i={1,2}$).
Then the total loss $L$ is used to update the parameters of the encoder $E_{\theta_{1}}$ as follows:
\begin{equation}
\label{eq4}
\theta_{1} \leftarrow \mathrm{Opt}(\theta_{1}, \nabla_{\theta_{1}}L, \alpha),
\end{equation}
where $\mathrm{Opt}$ and $\alpha$ represent the optimizer and learning rate, respectively.
The weights of $E_{\theta_{2}}$ are an exponential moving average of the weights of $E_{\theta_{1}}$ and are updated as follows:
\begin{equation}
\label{eq5}
\theta_{2} \leftarrow \tau\theta_{2} + (1-\tau)\theta_{1},
\end{equation}
where $\tau$ represents the moving average degree.
The gradient is not backpropagated through the encoder $E_{\theta_{2}}$ for stable training~\cite{chen2021exploring}.
Thus, we can learn discriminative representations from CXR images using the transfer learning from natural images and SSL on CXR images.
After the SSL process, we fine-tuned the labeled CXR images for COVID-19 detection using the encoder $E_{\theta_{1}}$.
\begin{table}[t!]
    \small
	\begin{center}
		\caption{Hyperparameters of the proposed method.}
		\label{tab1}
		\begin{tabular}[t]{lc}
			\hline
			Hyperparameter  & Value \\
			\hline
            SSL epoch & 40 \\
            Fine-tuning epoch & 30 \\ 
            Batch size    & 256 \\
		    Learning rate ($\alpha$) & 0.03 \\
		    Momentum      & 0.9 \\
		    Weight decay  & 0.0004 \\
			Moving average ($\tau$) & 0.996 \\
			MLP hidden size & 4096 \\
			Projection size & 256 \\
			View size      & 128 \\
			\hline
		\end{tabular}
	\end{center}
\end{table}
\begin{table*}[t]
    \centering
    \caption{COVID-19 detection results compared with different methods.}
    \label{tab2}
    \begin{tabular}{lccccc}
    \hline
    Method & Sen & Spe & HM & AUC & Acc \\\hline
    Ours (Transfer + SSL)
    & \bfseries{0.972$\pm$0.003} & \bfseries{0.997$\pm$0.001} & \bfseries{0.985$\pm$0.001} & \bfseries{0.999$\pm$0.000} & \bfseries{0.953$\pm$0.001} \\
    Transfer
    & 0.944$\pm$0.004 & 0.994$\pm$0.001 & 0.968$\pm$0.002 & 0.997$\pm$0.000 & 0.936$\pm$0.001 \\
    Cross~\cite{li2021self}
    & 0.923$\pm$0.005 & 0.991$\pm$0.001 & 0.955$\pm$0.002 & 0.995$\pm$0.000 & 0.908$\pm$0.001 \\
    BYOL~\cite{grill2020bootstrap}
    & 0.895$\pm$0.005 & 0.987$\pm$0.001 & 0.939$\pm$0.003 & 0.991$\pm$0.000 & 0.894$\pm$0.001 \\
    SimSiam~\cite{chen2021exploring}
    & 0.794$\pm$0.013 & 0.972$\pm$0.002 & 0.874$\pm$0.007 & 0.972$\pm$0.000 & 0.849$\pm$0.001 \\
    SimCLR~\cite{chen2020simple}
    & 0.778$\pm$0.006  & 0.965$\pm$0.002 & 0.862$\pm$0.003 & 0.996$\pm$0.000 & 0.876$\pm$0.001 \\
    PIRL-Jigsaw~\cite{misra2020self}
    & 0.685$\pm$0.014 & 0.973$\pm$0.003 & 0.804$\pm$0.009 & 0.954$\pm$0.000 & 0.821$\pm$0.001 \\
    PIRL-Rotation~\cite{misra2020self}
    & 0.760$\pm$0.009 & 0.962$\pm$0.002 & 0.849$\pm$0.005 & 0.960$\pm$0.001 & 0.817$\pm$0.001 \\
    From Scratch
    & 0.665$\pm$0.013 & 0.954$\pm$0.003 & 0.783$\pm$0.008 & 0.935$\pm$0.001 & 0.774$\pm$0.002 \\\hline
    \end{tabular}
\end{table*}
\begin{figure*}[t]
        \centering
        \subfigure[Ours]{
        \centering
        \includegraphics[width=5.3cm]{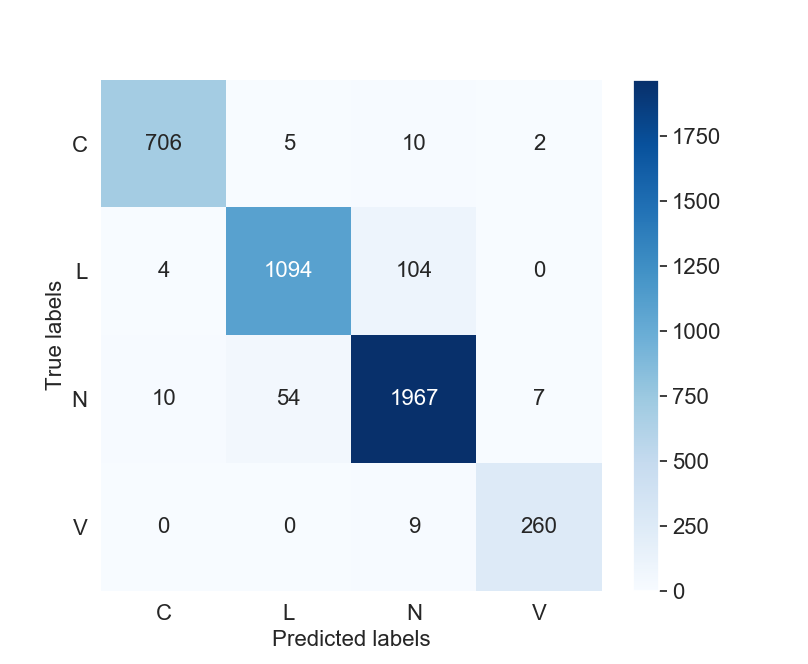}
        }
        \subfigure[Transfer]{
        \centering
        \includegraphics[width=5.3cm]{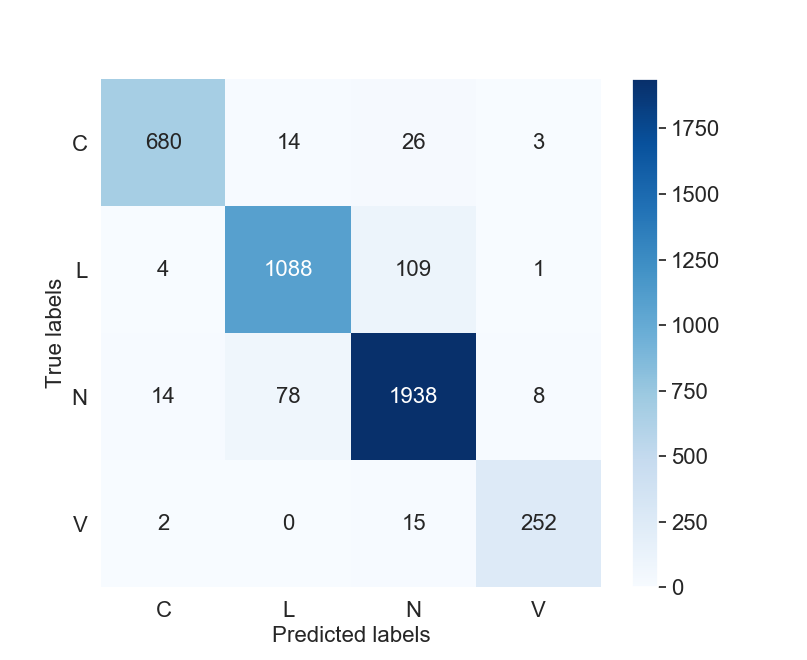}
        }
        \subfigure[Cross]{
        \centering
        \includegraphics[width=5.3cm]{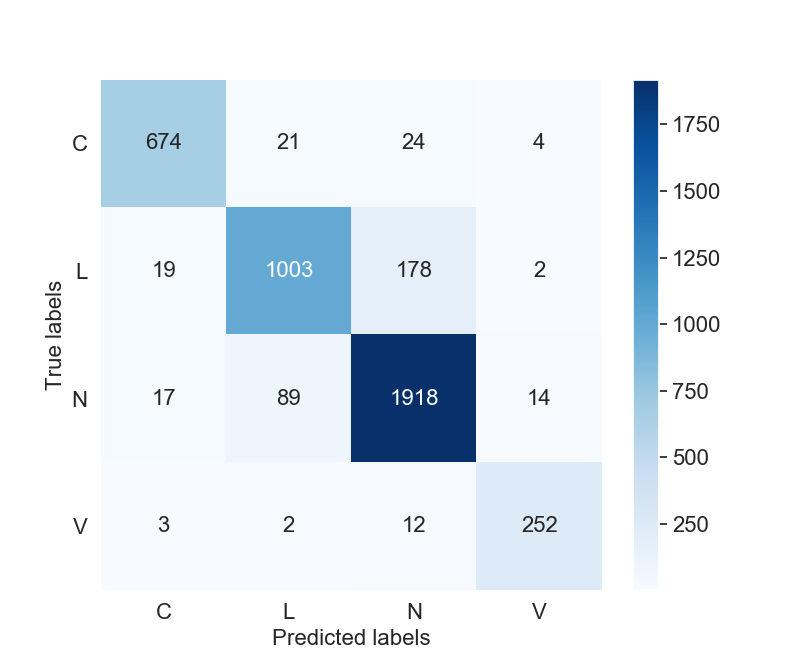}
        }
        \subfigure[BYOL]{
        \centering
        \includegraphics[width=5.3cm]{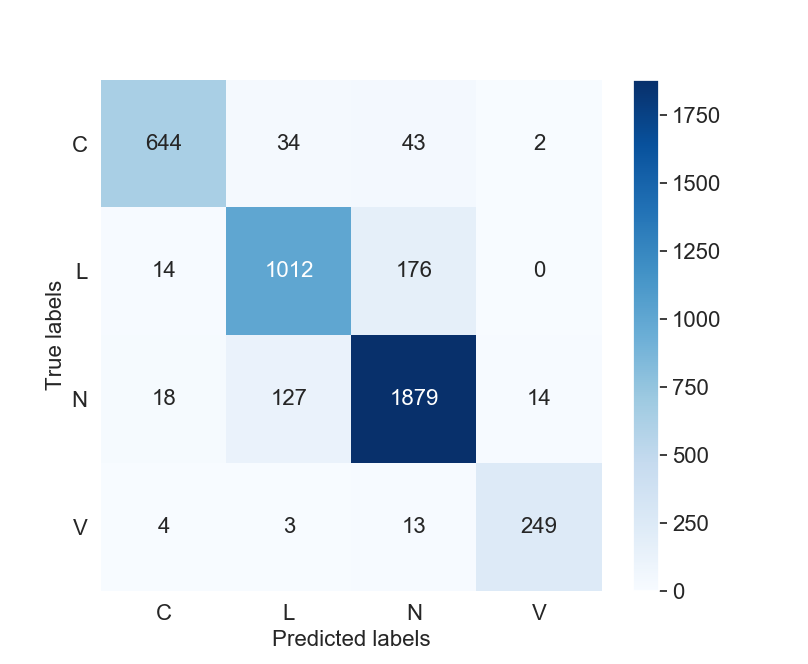}
        }
        \subfigure[SimSiam]{
        \centering
        \includegraphics[width=5.3cm]{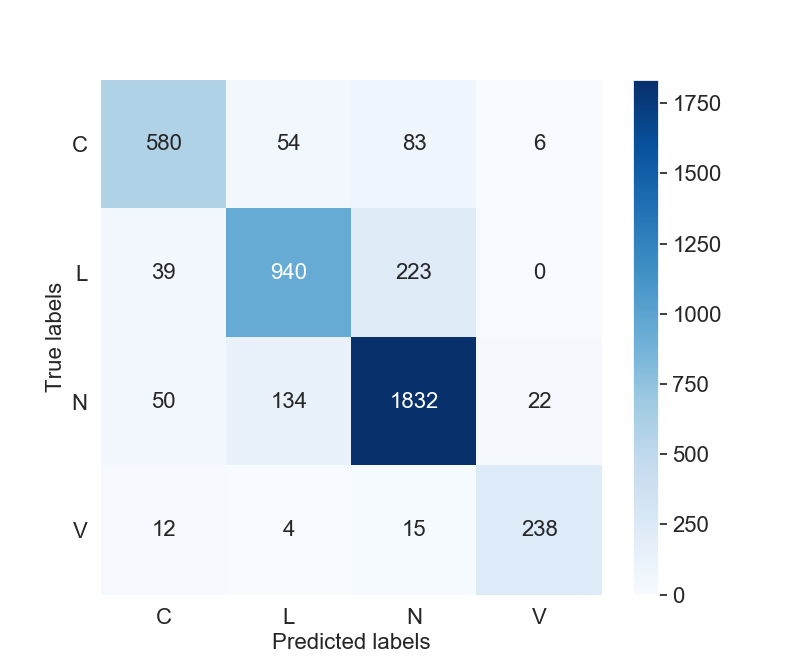}
        }
        \subfigure[SimCLR]{
        \centering
        \includegraphics[width=5.3cm]{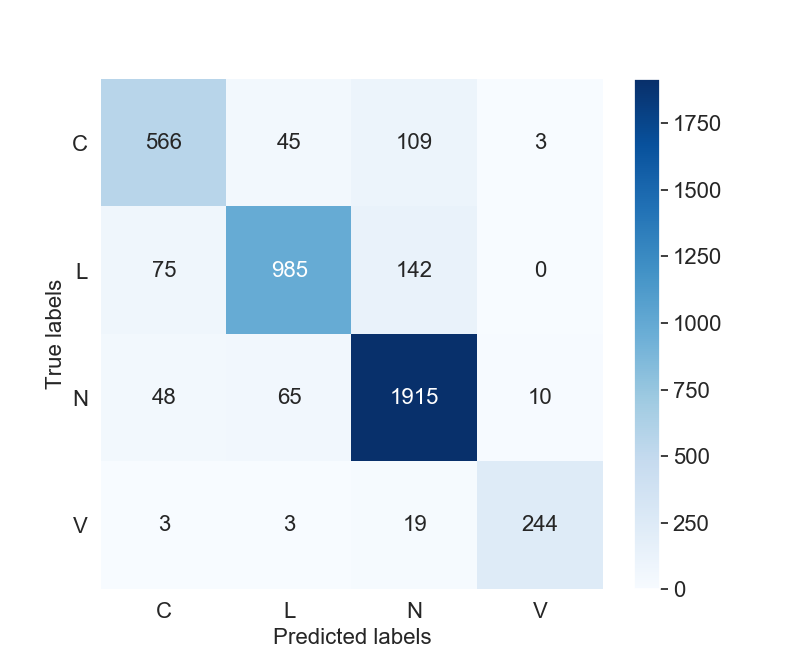}
        }
        \subfigure[PIRL-Jigsaw]{
        \centering
        \includegraphics[width=5.3cm]{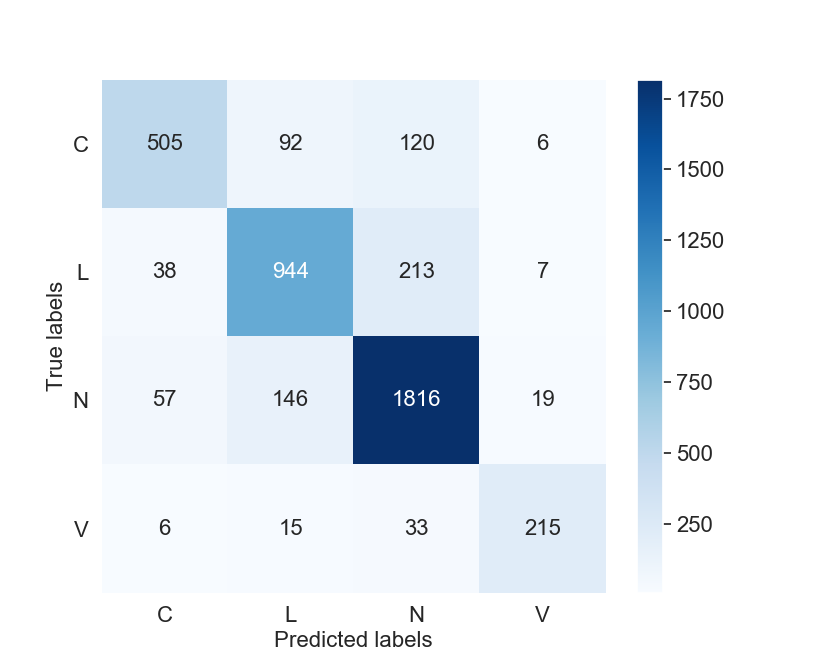}
        }
        \subfigure[PIRL-Rotation]{
        \centering
        \includegraphics[width=5.3cm]{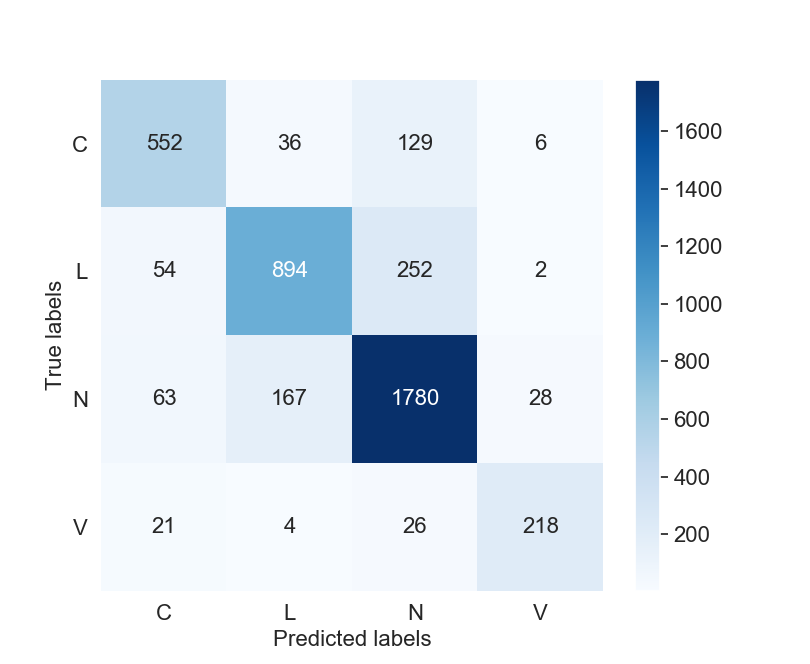}
        }
        \subfigure[From Scratch]{
        \centering
        \includegraphics[width=5.3cm]{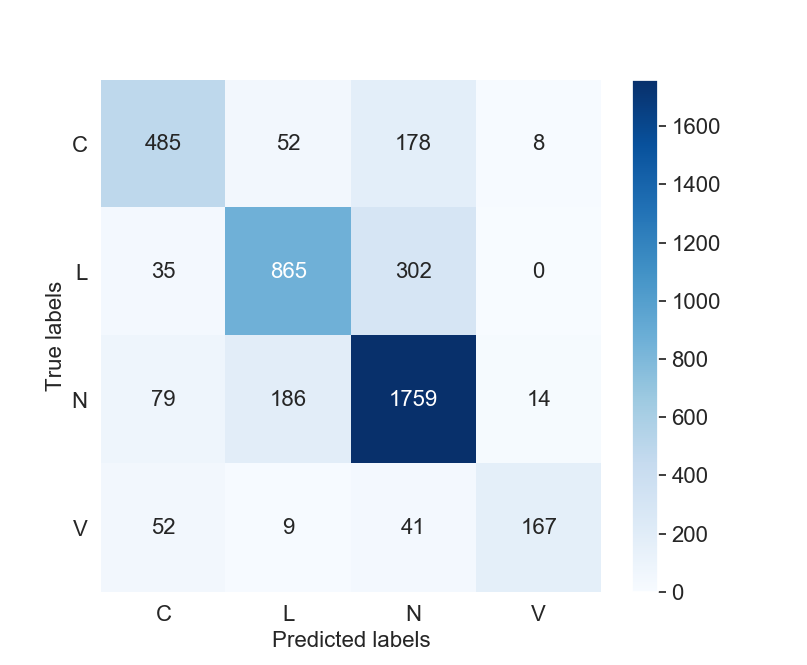}
        }
        \caption{Confusion matrix of our method (a) and other comparative methods (b)--(i).}
        \label{fig2}
\end{figure*}
\begin{figure*}[t]
        \centering
        \subfigure[]{
        \centering
        \includegraphics[width=8.3cm]{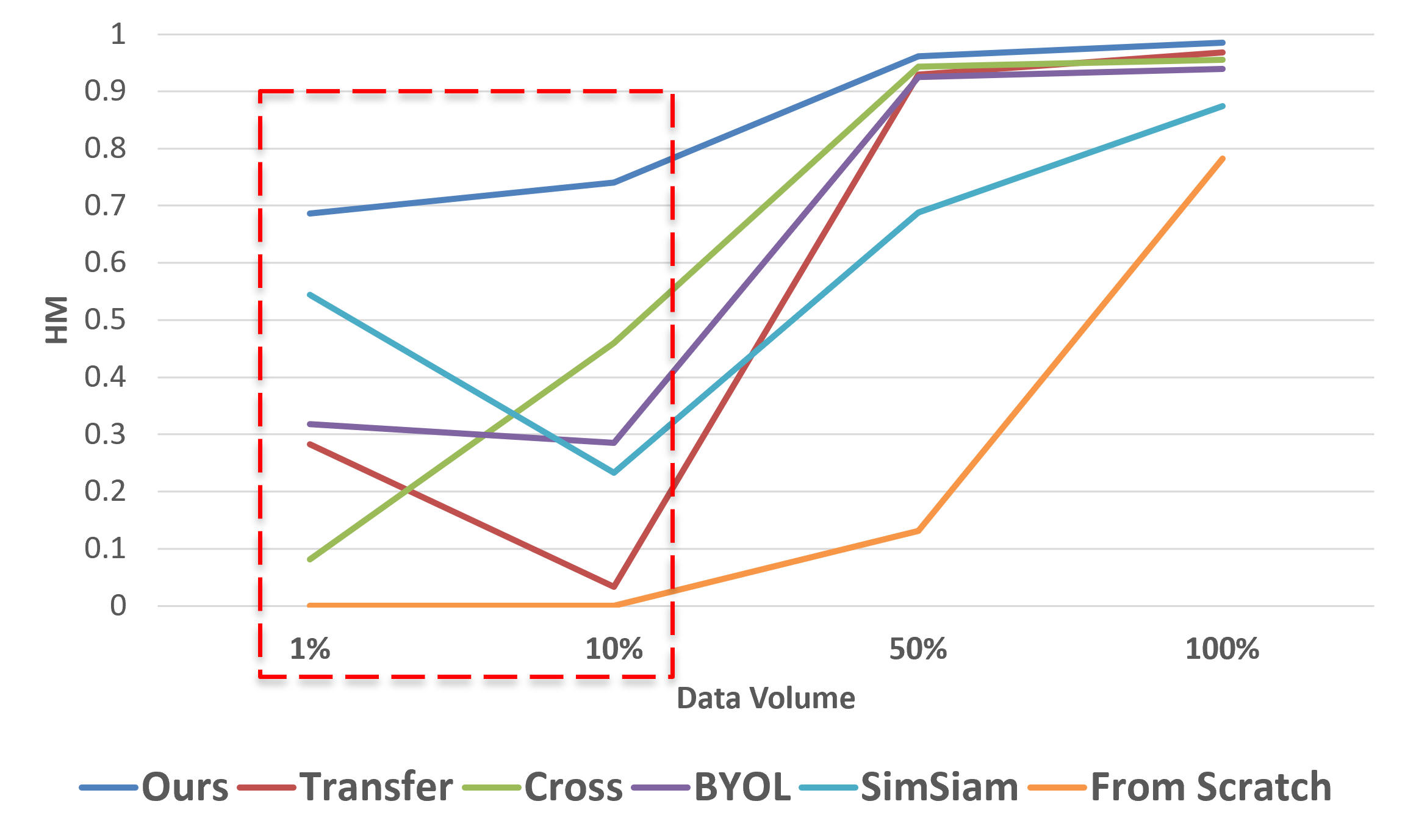}
        }
        \subfigure[]{
        \centering
        \includegraphics[width=8.3cm]{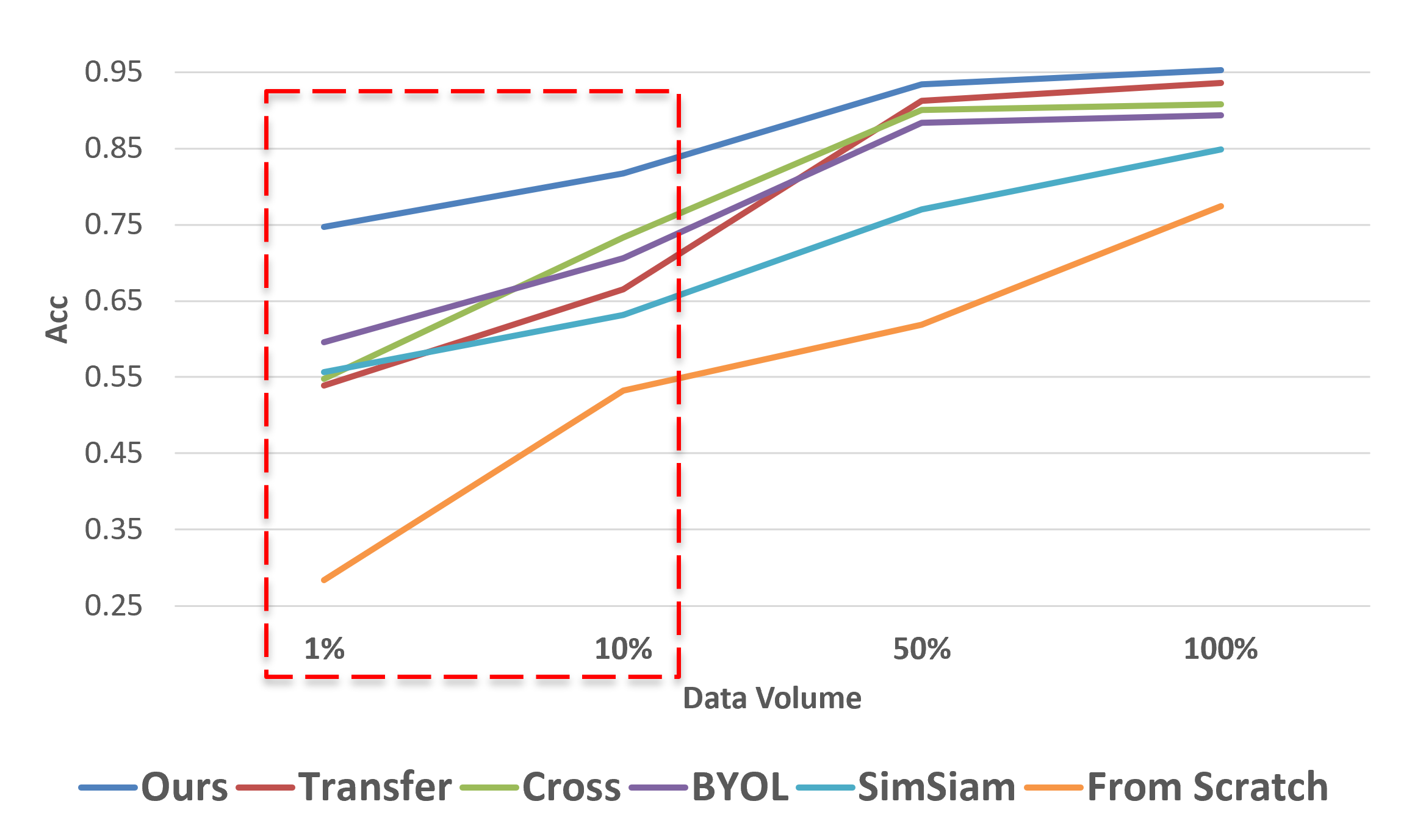}
        }
        \caption{COVID-19 detection results in different fine-tuning data volumes: (a) HM and (b) Acc.}
        \label{fig3}
\end{figure*}
\begin{figure*}[t]
        \centering
        \includegraphics[width=18cm]{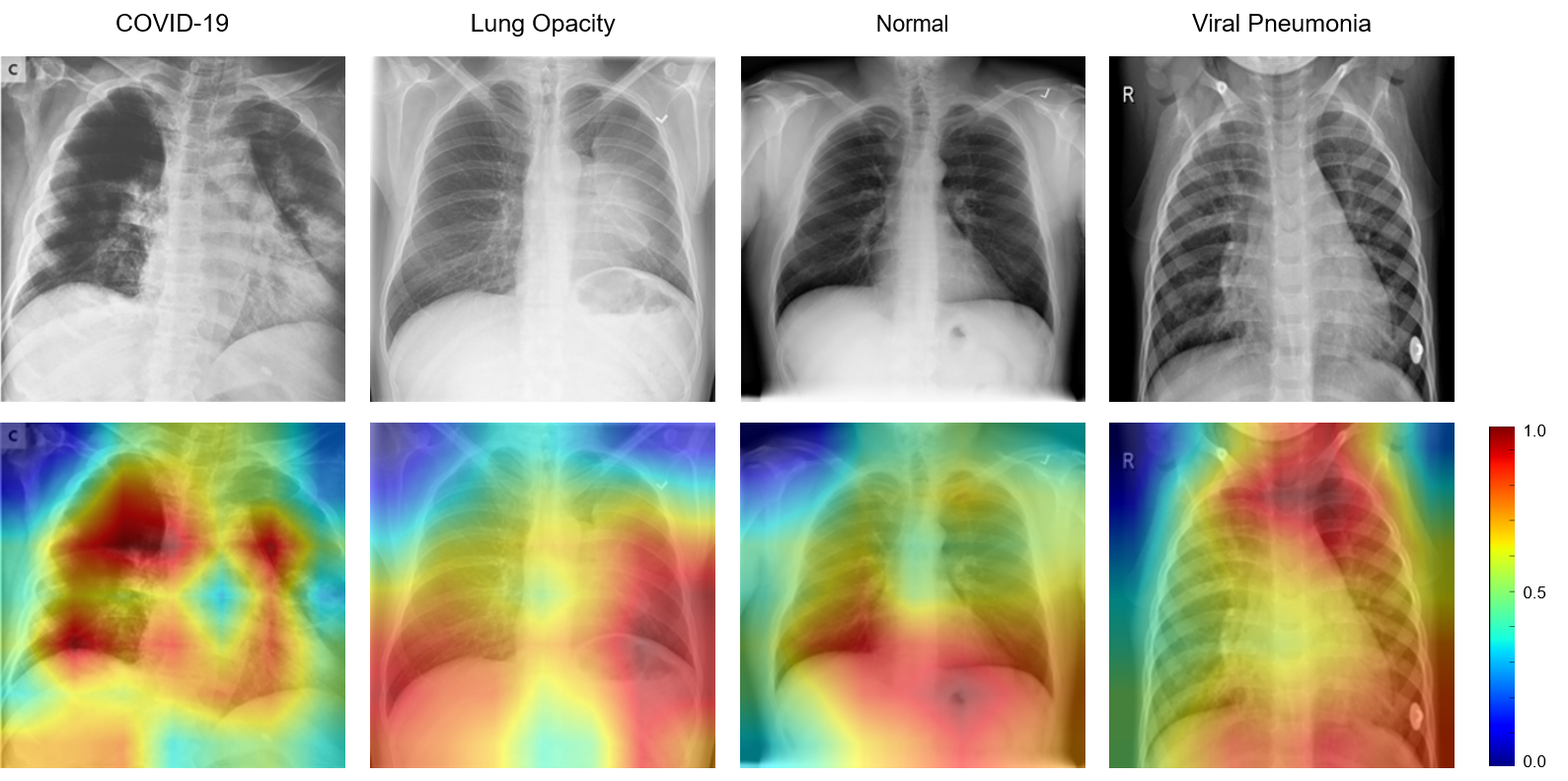}
        \caption{Grad-CAM++ visual explanations of the proposed method. Red represents high-attention and blue represents low-attention.}
        \label{fig4}
\end{figure*}
\section{Methodology}
\label{sec3}
\subsection{Dataset and settings}
\label{sec3-A}
We used the largest open COVID-19 CXR dataset in this study~\cite{rahman2021exploring}.
The dataset has 21,165 CXR images in four categories, all CXR images are 299 $\times$ 299 pixels and in png format.
In this dataset, 3,616 positive COVID-19 CXR images are collected from public datasets, publications, and websites.
More details about the used dataset can be found on the official website.\footnote{https://www.kaggle.com/datasets/tawsifurrahman/covid19-radiography-database}
One example of each category in the dataset is shown in Fig.~\ref{fig4}.
80\% of the dataset were used as the training set and the remaining 20\% were used as the test set.
Sensitivity (Sen), specificity (Spe), the harmonic mean (HM) of Sen and Spe (Eqs.~(\ref{eq6})-(\ref{eq8})), the area under the ROC curve (AUC), and the accuracy (Acc) were used as evaluation indexes~\cite{li2022self}.
\begin{equation}
\label{eq6}
\mathrm{Sen} = \frac{\mathrm{TP}}{\mathrm{TP + FN}},
\end{equation}
\begin{equation}
\label{eq7}
\mathrm{Spe} = \frac{\mathrm{TN}}{\mathrm{TN + FP}},
\end{equation}
\begin{equation}
\label{eq8}
\mathrm{HM} = \frac{\mathrm{2 \times Sen \times Spe}}{\mathrm{Sen + Spe}},
\end{equation}
where TP, TN, FP, and FN represent the number of true positive, true negative, false positive, and false negative, respectively. 
For calculating Sen, Spe, HM, and AUC, the COVID-19 category is taken as positive, and the other categories are taken as negative.
We used ResNet50 encoder and stochastic gradient descent optimizer.
Hyperparameters of the proposed method (e.g., $\alpha$ in Eq.~(\ref{eq4}) and $\tau$ in Eq.~(\ref{eq5})) are shown in Table~\ref{tab1}.
All 21,165 CXR images without label information in the training set were used for self-supervised pre-training.
All experiments were conducted using the PyTorch framework with an NVIDIA Tesla P100 GPU with 16G memory. 
The training time of the self-supervised pre-training and supervised fine-tuning is about 47 and 35 minutes, respectively.
\par
We used several SSL methods such as Cross~\cite{li2021self}, BYOL~\cite{grill2020bootstrap}, SimSiam~\cite{chen2021exploring}, PIRL-Jigsaw~\cite{misra2020self}, and SimCLR~\cite{chen2020simple}, transfer learning (using ImageNet~\cite{deng2009imagenet} pre-trained weights), and training from scratch as comparative methods.
Note that PIRL-Jigsaw and PIRL-Rotation are based on jigsaw and rotation pretext tasks, respectively.
Additionally, Additionally, we compared six pretrained DCNNs (ResNet18, ResNet50, ResNet101, InceptionV3~\cite{szegedy2016rethinking}, DenseNet201, and CheXNet~\cite{rajpurkar2017chexnet}) with the proposed method.
To verify the effectiveness of our method even when using a few labeled data, we randomly selected data from the training set (1\%, 10\%, and 50\%).
Note that the ratio selected in each category is the same.
\subsection{Experimental results}
\label{sec3-B}
COVID-19 detection results compared with different methods are shown in Table~\ref{tab2} and Fig.~\ref{fig2}.
The results show the average and variance of the last 10 fine-tuning epochs.
From Table~\ref{tab2}, our method drastically outperformed other comparative methods and significantly improve COVID-19 detection performance compared with using SSL or transfer learning alone.
Specifically, when using all training data, transfer learning achieved HM, AUC, and Acc scores of 0.968, 0.997, and 0.936, respectively, and the best SSL method Cross achieved HM, AUC, and Acc scores of 0.955, 0.995, and 0.908, respectively.
However, when using both transfer learning and SSL, our method achieved HM, AUC, and Acc scores of 0.985, 0.999, and 0.953, respectively.
Figure~\ref{fig2} shows the confusion matrix of our method (a) and other comparative methods (b)--(i).
We can see that our method has great advantages in the recognition of all categories.
\par
Furthermore, Table~\ref{tab3} shows COVID-19 detection results of our method and different models reported in~\cite{rahman2021exploring}.
Specifically, the best model InceptionV3 achieved Sen of 0.935, Spe of 0.955, HM score of 0.945, and Acc of 0.935.
However, when using transfer learning and SSL, our method with ResNet50 achieved Sen of 0.972, Spe of 0.997, HM score of 0.985, and Acc of 0.953.
Experimental results show that the knowledge learned from natural images using transfer learning is beneficial for SSL on the CXR images and boosts representation learning for COVID-19 detection.
\begin{table}[t]
    \centering
    \caption{COVID-19 detection results of our method and different models reported in~\cite{rahman2021exploring}.}
    \label{tab3}
    \begin{tabular}{lcccc}
    \hline
    Method & Sen & Spe & HM & Acc \\\hline
    Ours
    & \bfseries{0.972} & \bfseries{0.997} & \bfseries{0.985} & \bfseries{0.953} \\
    ResNet18
    & 0.934 & 0.955 & 0.944 & 0.934 \\
    ResNet50
    & 0.930 & 0.955 & 0.942  & 0.930 \\
    ResNet101
    & 0.930 & 0.951 & 0.940 & 0.930 \\
    InceptionV3
    & 0.935 & 0.955 & 0.945 & 0.935 \\
    DenseNet201
    & 0.927 & 0.954 & 0.940 & 0.927 \\
    CheXNet
    & 0.932 & 0.955 & 0.943 & 0.932 \\
    \hline
    \end{tabular}
\end{table}
\par
COVID-19 detection results in different training data volumes are shown in Fig.~\ref{fig3}.
The results contain the average of the last 10 fine-tuning epochs.
From Fig.~\ref{fig3}, our method significantly improved COVID-19 detection in small data volumes, such as 1\% and 10\% of the training set (169 and 1,693 images) compared with other methods, and achieved promising detection performance even for 50\% of the training set.
Examples of CXR images and their Grad-CAM++~\cite{chattopadhay2018grad} visual explanations of the proposed method are shown in Fig.~\ref{fig4}, where the highlight regions are used for decision-making.
These visualization results increase the confidence and reliability of the proposed method, by confirming the accuracy of the decision-making on the relevant region of the CXR images~\cite{brunese2020explainable}.
\section{Discussion}
\label{sec4}
Considering several patients screened due to COVID-19 pandemic, the use of deep learning for computer-aided detection has strong potential in assisting clinical workflow efficiency and reducing the incidence of infections among radiologists and healthcare providers~\cite{bertolini2021high, shalbaf2021automated}.
Here, we proposed a new learning scheme called self-supervised transfer learning for COVID-19 detection using CXR images.
Our findings show that the proposed method boosts COVID-19 detection on the largest open COVID-19 CXR dataset.
Especially, when using a small amount of labeled training data for the final fine-tuning, our method drastically outperformed other methods.
\par
Due to the completely different infection status, the number of medical resources, and data sharing policies of COVID-19 in different countries and cities, there is the likelihood of limited labeled training data~\cite{peiffer2020machine}.
Nonetheless, the proposed method can still be applied to this case for high-performance COVID-19 detection.
Our work sprouts from clinical needs and the proposed method can be applied to other diseases detection not only COVID-19.
Although the experimental results are promising, the proposed method should be evaluated on other COVID-19 datasets or datasets of different diseases for any potential bias.
Moreover, our previous study~\cite{li2020soft, li2022compressed} can improve the effectiveness and security of medical data sharing among different medical facilities, which fits well with the proposed self-supervised transfer learning and is expected to be applied in clinical.
\par
Our method also has limitations.
For example, our method has three stages and hence is not end-to-end.
The encoder should be saved in the pre-training stage and continue to update the parameters in the later stages, which is more complicated in operation.
In our experiments, we want to explore the impact and robustness of the initial parameters on different fine-tuning stages and data volumes. 
Hence, we use the average of the last 10 fine-tuning epochs to test the performance. 
Also, since we evaluate model performance in different subsets of settings (i.e., 1\%, 10\%, and 50\%), it is expensive to perform N-fold cross-validation in all settings.
However, N-fold cross-validation is a more common method, which we will consider in our future work.
\section{Conclusion}
\label{sec5}
A new learning scheme called self-supervised transfer learning for detecting COVID-19 from CXR images has been proposed in this paper.
We showed that the knowledge learned from natural images with transfer learning benefits the SSL on the CXR images and boosts the representation learning performance for COVID-19 detection. 
Our method can learn discriminative representations from CXR images without manually annotated labels.
Experimental results showed that our method achieved promising results on the COVID-19 CXR dataset.
Furthermore, our method can help reduce the incidence of infections among radiologists and healthcare providers.
\section*{Ethical approval}
No ethics approval is required.
\section*{Declaration of competing interest}
None declared.
\section*{Acknowledgments}
This study was supported in part by AMED Grant Number JP21zf0127004, the Hokkaido University-Hitachi Collaborative Education and Research Support Program, and the MEXT Doctoral program for Data-Related InnoVation Expert Hokkaido University (D-DRIVE-HU) program. This study was conducted on the Data Science Computing System of Education and Research Center for Mathematical and Data Science, Hokkaido University.
\bibliographystyle{spmpsci}
\bibliography{refs}

\end{document}